\documentclass[a4paper,11pt]{article}
\usepackage{pos}
\usepackage[backend=bibtex,url,doi=false,hyperref,style=numeric,sorting=none,bibencoding=utf8,giveninits=true,maxnames=2]{biblatex}
\usepackage{dcolumn}
\usepackage{siunitx}
\usepackage{soul}
\usepackage[absolute,overlay]{textpos}
\usepackage{staves}
\usepackage{xcolor}
\usepackage{changepage}
\usepackage{etex}
\usepackage{tikz}
\usepackage{pgflibrarysnakes}
\usepackage{url}
\usepackage{graphicx}
\usepackage{amsmath}
\usepackage{bm}
\usepackage{booktabs}
\usepackage{textcomp}
\usetikzlibrary{decorations.markings}
\usetikzlibrary{svg.path}
\usepackage{enumitem}
\usepackage{pgfplots}
\usepackage{caption}
\usepackage{tikz-feynman}

\newcommand{\abs}[1]{\lvert#1\rvert}
\newcommand{\eval}[1]{\langle#1\rangle}

\newcommand{\Nf}{N_\mathrm{f}}

\makeatletter
\DeclareFontFamily{OMX}{MnSymbolE}{}
\DeclareSymbolFont{MnLargeSymbols}{OMX}{MnSymbolE}{m}{n}
\SetSymbolFont{MnLargeSymbols}{bold}{OMX}{MnSymbolE}{b}{n}
\DeclareFontShape{OMX}{MnSymbolE}{m}{n}{
    <-6>  MnSymbolE5
   <6-7>  MnSymbolE6
   <7-8>  MnSymbolE7
   <8-9>  MnSymbolE8
   <9-10> MnSymbolE9
  <10-12> MnSymbolE10
  <12->   MnSymbolE12
}{}
\DeclareFontShape{OMX}{MnSymbolE}{b}{n}{
    <-6>  MnSymbolE-Bold5
   <6-7>  MnSymbolE-Bold6
   <7-8>  MnSymbolE-Bold7
   <8-9>  MnSymbolE-Bold8
   <9-10> MnSymbolE-Bold9
  <10-12> MnSymbolE-Bold10
  <12->   MnSymbolE-Bold12
}{}

\let\llangle\@undefined
\let\rrangle\@undefined
\DeclareMathDelimiter{\llangle}{\mathopen}%
                     {MnLargeSymbols}{'164}{MnLargeSymbols}{'164}
\DeclareMathDelimiter{\rrangle}{\mathclose}%
                     {MnLargeSymbols}{'171}{MnLargeSymbols}{'171}
\makeatother

\AtEveryBibitem{\clearfield{title}}

\usetikzlibrary{decorations.markings}
\usetikzlibrary{patterns}
\usetikzlibrary{svg.path}
\usetikzlibrary{shapes,fit}
\usetikzlibrary{shapes.misc}

\usetikzlibrary{fit}
\tikzset{%
  highlight/.style={rectangle,fill=gray,
    fill opacity=0.5,thick,inner sep=0pt}
}
\tikzset{cross/.style={cross out, draw=black, thick, minimum size=2*(#1-\pgflinewidth), inner sep=0pt, outer sep=0pt},
cross/.default={3pt}}

\title{Vacuum correlators at short distances from lattice QCD}

\author*[a]{Tim Harris}
\author[b]{Marco C\`e}
\author[c,d,e]{Harvey B. Meyer}
\author[c]{Arianna Toniato}
\author[c]{Csaba T\"or\"ok}

\affiliation[a]{School of Physics and Astronomy, University of Edinburgh, EH9 3JZ, UK}
\affiliation[b]{Theoretical Physics Department, CERN, CH-1211 Geneva 23, Switzerland}
\affiliation[c]{PRISMA$^+$ Cluster of Excellence \& Institut f\"ur
Kernphysik, Johannes Gutenberg-Universit\"at Mainz,\\
Saarstr.\ 21, 55122 Mainz, Germany}
\affiliation[d]{Helmholtz Institut Mainz, Johannes
Gutenberg-Universit\"at Mainz,
Saarstr.\ 21, 55122 Mainz, Germany}
\affiliation[e]{GSI Helmholtzzentrum f\"ur Schwerionenforschung,
Planckstra\ss{}e 1, 64291, Darmstadt, Germany}

\emailAdd{tharris@ed.ac.uk}

\abstract{
We propose a method to help control cutoff effects in the short-distance
contribution to integrated correlation functions, such as the hadronic vacuum
polarization (HVP), using the corresponding screening correlators computed at finite
temperature.
The strategy is investigated with Wilson fermions at leading order, which
reveals a logarithmically-enhanced lattice artifact in the short-distance
contribution, whose coefficient is determined at this order.
We then perform a numerical study with $N_\mathrm{f}=2$ O($a$)-improved Wilson
fermions and a temperature $T\approx250$~MeV, with lattice spacings down to
$a\approx0.03$~fm, which suggests good control can be achieved on the short-distance
contribution to the HVP and the Adler function at
large virtuality.
Finally, we put forward a scheme to compute the complete HVP
function at arbitrarily large virtualities using a step-scaling
in the temperature.
\begin{flushright}
    CERN-TH-2021-189
\end{flushright}
}

\FullConference{%
 The 38th International Symposium on Lattice Field Theory, LATTICE2021
  26th-30th July, 2021
  Zoom/Gather@Massachusetts Institute of Technology
}

\begin{document}
\maketitle

\section{Introduction}
\label{sec:intro}

\begin{figure}[t]
    \centering
    \includegraphics[scale=0.8]{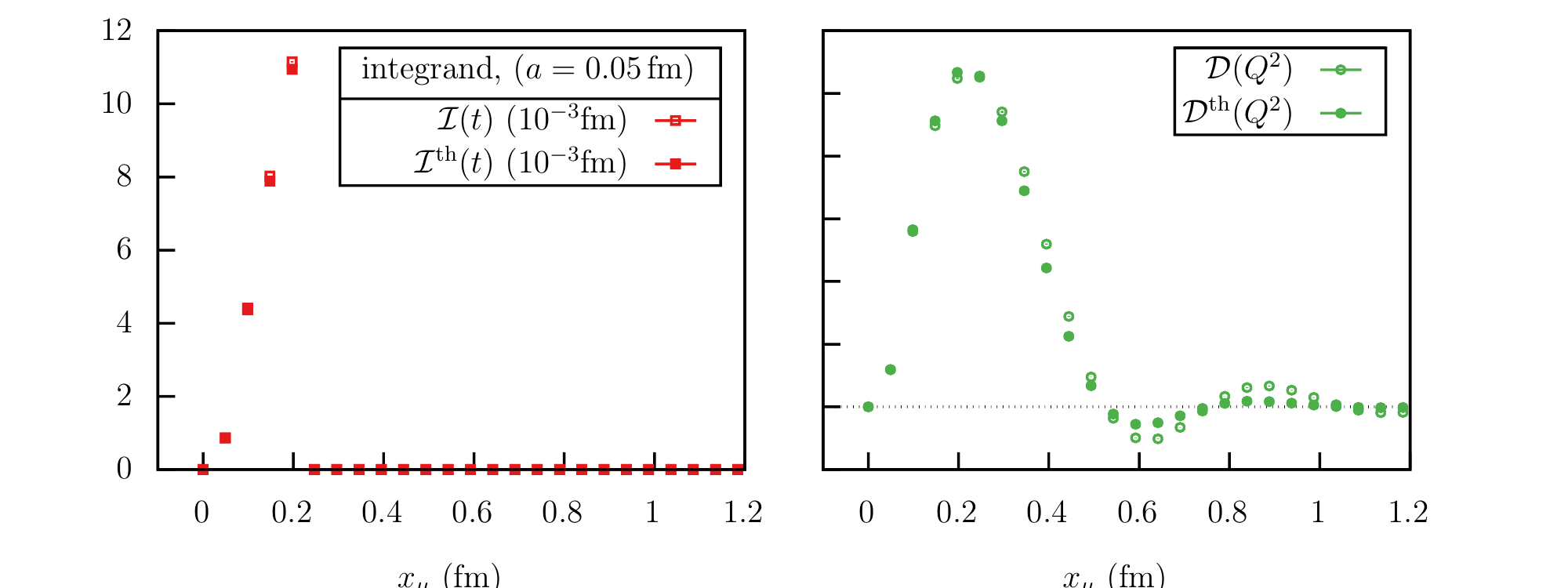}
    \caption{The integrand of the fourth moment truncated at
        $t\sim0.2\,\mathrm{fm}$ (left) and the Adler function with
        $Q\sim2.4\,\mathrm{GeV}$ (right) at finite lattice spacing
        $a\sim0.05\,\mathrm{fm}$. The integrals are dominated by the
    contributions around $x_0\sim0.2\,\mathrm{fm}$. The filled symbols
    depict the thermal observable, while the open symbols represent the
    corresponding vacuum contribution.}
    \label{fig:integrand}
\end{figure}

Correlation functions evaluated at small distances are susceptible to large
cutoff effects, which are crucial to control to obtain reliable estimates for
many interesting physical observables.
In many contexts, such short-distance contributions arise naturally when
correlators, such as that of the electromagnetic current,
\begin{align}
    \label{eq:curr_corr}
    G(x_0) &= -\int\mathrm{d}^3x \,\eval{J_1(x)J_1(0)},\qquad 
    J_\mu=\sum_fQ_f\bar\psi_f\gamma_\mu\psi_f,
\end{align}
are integrated over all separations $x_0$ with a known weight function.
Examples are given by the short-distance HVP
contribution to the lepton anomalous magnetic moment,
\begin{align}
    \label{eq:sdhvp}
    I(t) &= \int_0^t\mathrm{d}x_0 \,x_0^4 G(x_0),
\end{align}
(which to a good approximation is independent of the lepton mass for
small enough $t$), or the Adler function at large virtualities
\begin{align}
    D(Q^2) = \frac{12\pi^2}{Q^2}\int_0^\infty\mathrm{d}x_0
    \Large[2-2\cos(Qx_0)-Qx_0\sin(Qx_0)\Large]G(x_0).
\end{align}
The integrands of these quantities are shown in figure~\ref{fig:integrand} on
the lattice with $\Nf=2$ Wilson fermions as described in
section~\ref{sec:nf2}, which illustrates they are dominated by the region
around $x_0\sim0.2\,\mathrm{fm}$ for the choices $t\sim0.2\,\mathrm{fm}$
and $Q\sim 2.4\,\mathrm{GeV}$.
In the following we concentrate on the first observable, the short-distance
HVP contribution to the lepton magnetic moment or
truncated fourth moment of the correlator, but the full results for the Adler
can be found in ref.~\cite{Ce:2021xgd}.

Power-counting suggests that the cutoff effects for the lattice correlator
$\mathcal G$ are enhanced at small distances~\cite{DellaMorte:2008xb}, where,
in the massless and and for $x_0\Lambda\ll1$, the only relevant scale is given
by the distance $x_0$, and we have
\begin{align}
    \mathcal G(x_0,a) &= G(x_0) + \mathrm{const.}({a}/{x_0})^2 G(x_0) +
    \ldots,
    \label{eq:cutoff}
\end{align}
assuming full $\mathrm{O}(a)$-improvement and ignoring the dependence of the running coupling on the cutoff scale.
As we show in the following section, this enhancement of the cutoff effects in
the correlation function leads to a logarithmic enhancement of the leading
$a^2$ discretization effects in the short-distance HVP contribution to the lepton magnetic moment already at leading
order with free fermions.

\paragraph{Thermal improvement}

\begin{figure}[t]
    \centering
    \scalebox{0.8}{\begin{tikzpicture}[scale=0.5]
    \begin{scope}[rotate=-90]
        \tikzset{dstyle/.style={shape=circle,fill=black,scale=0.3}}
        \tikzset{ostyle/.style={shape=rectangle,draw,fill=white,scale=0.5}}
        \fill[fill=white] (-4,-1.5) rectangle (-3,1);
        \foreach \x in {0,...,5}
        \foreach \y in {0,...,11}
        {
            \node[dstyle] (\x-\y) at (\x,\y) {};
        }

        \foreach \x in {0,...,10}
        \foreach \y in {28,...,34}
        {
            \node[dstyle] (\x-\y) at (\x/2,\y/2) {};
        }
        \node[draw,very thick,dotted,fit=(0-0) (5-11)] {};
        \node[draw,very thick,dotted,fit=(0-28) (10-34)] {};
        \draw[<->,very thick]  (-1,0)  -- (-1,11) node[midway,above] {{\Large $L_0$}};
        \draw[<->,very thick]  (0,-1)  -- (5,-1) node[midway,left] {{\Large $L$}};
        \draw[<->,very thick]  (-1,14)  -- (-1,17) node[midway,above] {{\Large $1/T$}};
        \draw[very thick,blue] (4,1) .. controls (4.5,2) and (4.5,3) .. (4,4);
        \draw[very thick,blue] (4,1) .. controls (3.5,2) and (3.5,3) .. (4,4);
        \draw[very thick,red] (5,15) .. controls (4,14.5) and (3,14.5) ..  (2,15);
        \draw[very thick,red] (5,15) .. controls (4,15.5) and (3,15.5) ..  (2,15);
        \draw[<->,very thick] (3,20) node[above] {\Large $\hat 3$} -- (5,20)
        -- (5,22) node[right] {\Large $\hat 0$};
    \end{scope}
\end{tikzpicture}%
}
    \caption{Illustration of the geometry of the vacuum lattices (left) and
    thermal lattices (right) with smaller temporal size but finer lattice
    spacing. Note also the correlators in the thermal ensemble are computed in
    the orthogonal screening direction.}
    \label{fig:lattices}
\end{figure}
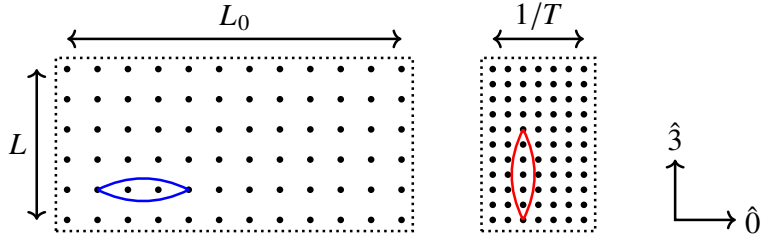

Due to the breakdown of the Symanzik effective theory at short distances,
it is desirable to seek an alternative approach to control cutoff effects in
integrated correlation functions.
In the following, we make use of the screening correlator analogous to
eq.~\eqref{eq:curr_corr}
\begin{align}
    \label{eq:scr_corr}
    G^\mathrm{th}(x_3) &= -\int\mathrm{d}x_0
    \mathrm{d}x_1\mathrm{d}x_2
    \,\eval{J_1(x)J_1(0)}_T
\end{align}
evaluated at temperature $T$, in a $T^{-1}\times L^3$ volume.
Compared with a vacuum ensemble of dimensions $L_0\times L^3$, the screening
correlator is relatively cheap to compute, as the volume is a factor
$L_0T$ smaller and one can reach smaller lattice spacings at a fixed cost,
which is illustrated in figure~\ref{fig:lattices}.
One can define the thermal analogues of the integrated quantities, such as the
fourth moment $I^\mathrm{th}$, by utilizing this screening quantity.
The operator-product expansion predicts that the leading $x_0^{-3}$
singularity of the correlators cancels when the thermal quantity is
subtracted, so that the relative difference in the integrated observable is
strongly suppressed with $tT$
\begin{align}
    \frac{I(t)-I^\mathrm{th}(t)}{I(t)} &= \mathrm O((tT)^3).
    \label{eq:diff}
\end{align}
Likewise, while the artifacts on the lattice estimator $\mathcal I$ are
$\mathrm O(a^2)$ up to logarithms in the $\mathrm O(a)$-improved theory, the
artifacts on the difference are also parametrically suppressed by $(tT)^3$.
This suggests writing a lattice estimator using the decomposition
\begin{align}
    \widehat{\mathcal I}(t) &= \underbrace{\mathcal I^\mathrm{th}}_{\mathrm
    O(a_\mathrm{th}^2)\,\textrm{artifacts}}
    + \underbrace{\Large[\mathcal
    I-\mathcal I^\mathrm{th}\Large]}_{\,\mathrm O(a^2(tT)^3)\textrm{
    artifacts}},
    \label{eq:imp_est}
\end{align}
where the first term on the right-hand side can be estimated using the thermal
simulations down to small lattice spacings $a_\mathrm{th}$, while the
remainder which depends on the costly vacuum ensembles can be estimated using
coarser lattices.
In particular, the size of the artifacts on the correction will be
parametrically smaller as long as the ratio of lattice spacings between the
thermal and vacuum lattices $a_\mathrm{th}/a$ does not fall below
$\sqrt{(tT)^3}$.

\section{Leading-order perturbative study}
\label{sec:lo}
In order to investigate the strategy, we computed the isovector vector
correlator at leading order in perturbation theory with $\Nf=2$ massless
Wilson fermions, using one local and one conserved current.
We expect the leading-order perturbative computation to capture the gross
features of the strategy given the short-distance nature of the observable.
In addition to providing a cross-check of the operator-product expansion at
leading order, this enabled a explicit computation of the lattice artifacts.

\paragraph{Logarithmic enhancement of lattice artifacts}
{The singular behaviour of the lattice artifacts observed in
eq.}~\eqref{eq:cutoff} {induces a logarithmic enhancement of the leading
$\mathrm O(a^2)$ lattice artifacts even at leading order in perturbation
theory.
This does not depend on the specific details of the lattice
discretization.}
For $\Nf=2$ Wilson fermions, and the isovector correlator with
$Q_\mathrm{u}=-Q_\mathrm{d}=\tfrac{1}{\sqrt 2}$, we find from an explicit
computation that 
\begin{align}
    \label{eq:lolog}
    \mathcal I &= I + c_{\mathcal I} a^2\log(1/a) + \mathrm{O}(a^2),\\
    c_\mathcal{I} &= \frac{7N_\mathrm{c}}{60\pi^2},
\end{align}
with a similar expression for the Adler function.
The leading-order logarithm observed here may be particularly severe as it
appears with the positive unit power, in constrast to the next-to-leading
order logarithms due to the dependence of the running coupling on the cutoff
scale~\cite{Husung:2019ytz}.
This enhancement, which is present in both the thermal and vacuum
quantities, illustrates the delicate nature of the continuum limit.

\begin{figure}[t]
    \begin{center}
        \includegraphics[scale=0.4]{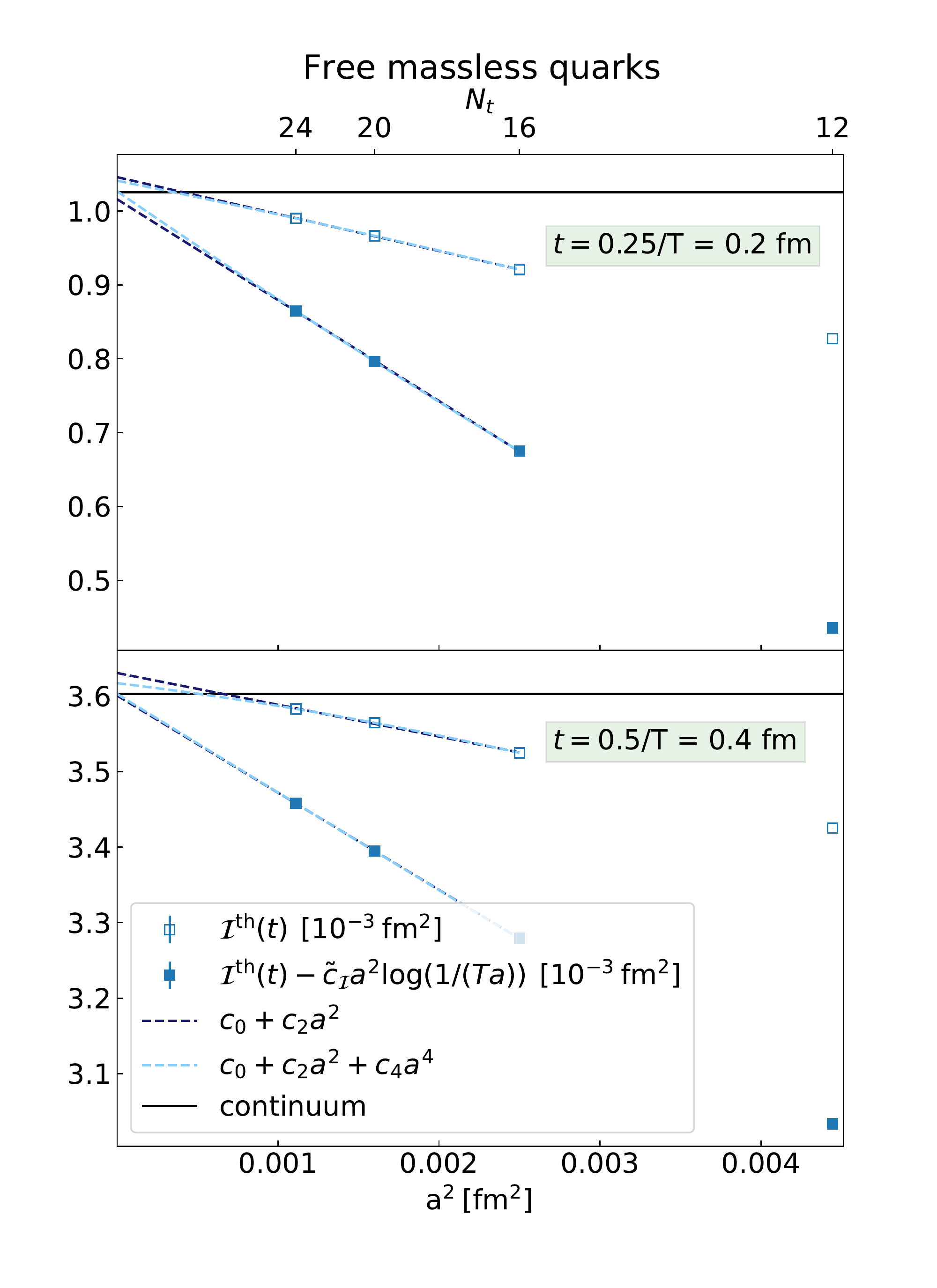}%
        \includegraphics[scale=0.4]{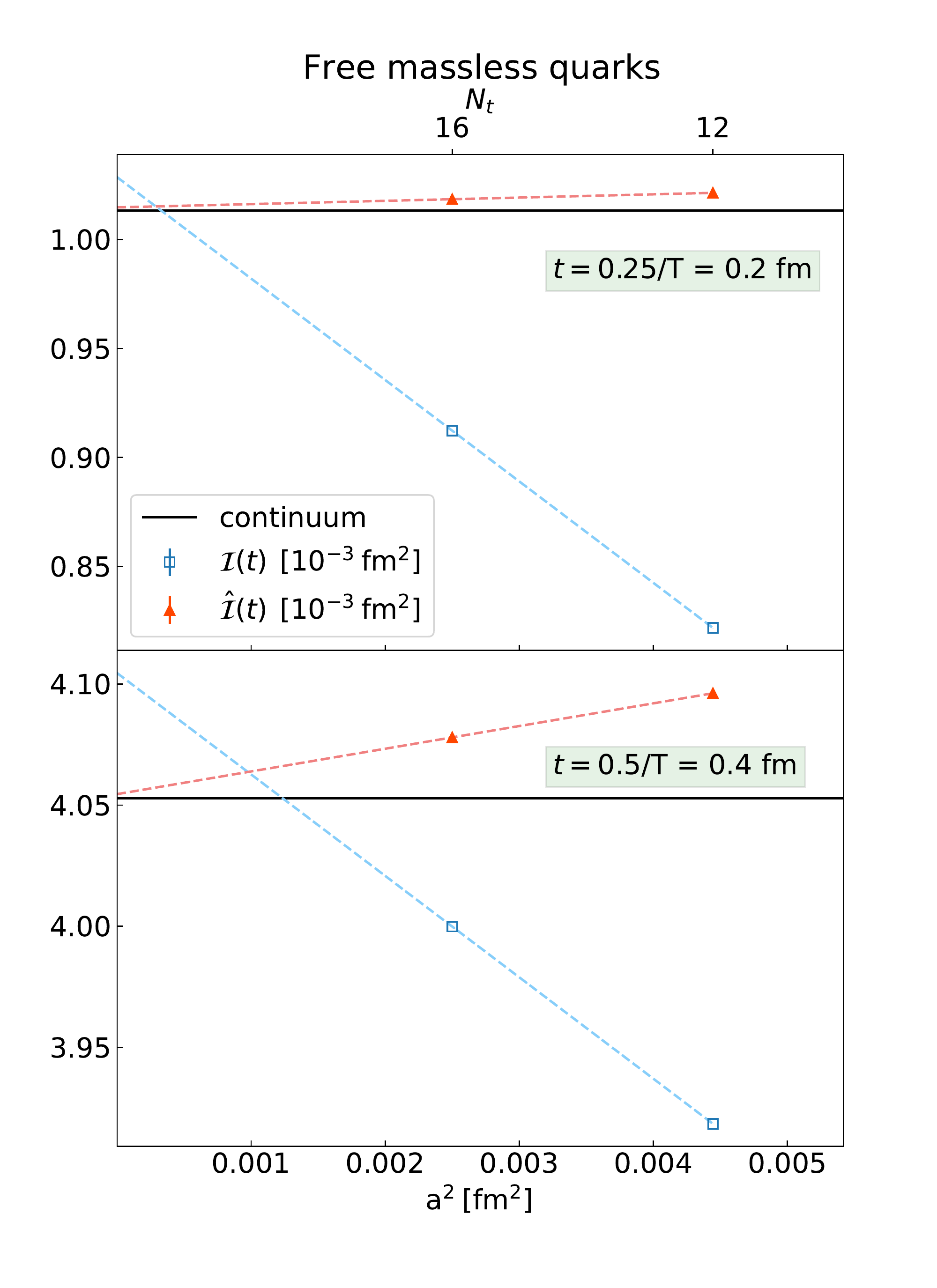}
    \end{center}
    \label{fig:lo}
    \caption{The continuum limit of the thermal observable (left) and vacuum
        observable (right). The top and bottom panels correspond to $tT=0.25$
        and $tT=0.5$ respectively. For the thermal case, two Ans\"atze are illustrated,
        and the data with and without the subtraction of the logarithmic artifact with
        the filled and open symbols respectively. In the vacuum case, the observable
        with (without) thermal subtraction is shown with red (blue) points.}
\end{figure}

\begin{table}[t]
    \centering
    \begin{tabular}[t]{c c c c c}
        \toprule
        $t$~(fm) & \multicolumn{2}{c}{$\abs{c_0 - I^{\mathrm{th}}(t)}/I^{\mathrm{th}}(t)$} & Ansatz  \\
        \midrule
        0.2 & 2\%   & 0.9\%    & $c_0 + c_2 a^2$           \\
            & 2\%   & 0.2 \%   & $c_0 + c_2 a^2 + c_4 a^4$ \\
        0.4 & 0.8\% & 0.06 \%  & $c_0 + c_2 a^2$           \\
            & 0.4\% & < 0.01\% & $c_0 + c_2 a^2 + c_4 a^4$ \\
        \midrule
        & plain & subtr. &   \\
        \bottomrule
    \end{tabular}
    \hspace{2em}
    \begin{tabular}[t]{c c c c}
    \toprule
    $t$~(fm) & \multicolumn{2}{c}{$\abs{c_0 - I(t)}/I(t)$} \\
    \midrule
    0.2 & 2\%          & 0.2\%              \\
    0.4 & 1\%          & 0.04 \%            \\
    \midrule
        & $\mathcal I$ & $\hat{\mathcal I}$  &  \\
    \bottomrule
    \end{tabular}
    \caption{The relative accuracy of the continuum limit in the leading-order
    case for the thermal observable (left) and the vacuum observable (right)
    using realistic lattice sizes and continuum extrapolations.  In the
    thermal case, either linear or quadratic fits in $a^2$ were used, with
    (``subtr.'') or without (``plain'') the subtraction of the logarithmic
    term. For the vacuum fits, the observable with and without the thermal
    subtraction is given.}
    \label{tab:lo}
\end{table}

To investigate the strategy of improvement, we computed the thermal lattice
observable with realistic thermal lattice sizes of $1/aT=12,16,20,24$, and the
thermal to vacuum correction using only coarser lattice sizes with
$1/aT=12,16$ in the thermal case.%
\footnote{These lattice sizes correspond to the state-of-the-art in
    non-perturbative simulations. Indeed, if the scale is set by the
    temperature to $T\approx250\,\mathrm{MeV}$ as we take as a reasonable choice
    later, then the smallest lattice spacing for the vacuum computation
correponds to $a\sim0.05\,\mathrm{fm}$ and $0.03\,\mathrm{fm}$ for the thermal
ones.}
For the thermal observable (left panel), the approach to the known continuum
limit is improved for both cases $tT=0.25$ (top) and $tT=0.5$ (bottom) if the
known logarithmic term is subtracted, which can be examined in detail in
table~\ref{tab:lo}.
For the vacuum case, the parametric reduction in the lattice artifacts
explained in the introduction is observed between the improved (red) and the
unimproved data (blue).
Without the thermal subtraction, an overestimate of this contribution is
observed at the per-cent level, whereas the case utilising the thermal
subtraction reaches easily sub-percent precision.

\section{Numerical results with \texorpdfstring{$\Nf=2$}{Nf=2} Wilson
fermions}
\label{sec:nf2}

\begin{table}[t]
    \begin{center}
        \begin{tabular}{c
                S[table-format=3]
                S[table-format=3]
                S[table-format=3]
                S[table-format=1.3]
                S[table-format=1.6]
                S[table-format=3]
            }
            \toprule
            \rule{0.5cm}{0pt}&\rule{0.5cm}{0pt}&\rule{0.5cm}{0pt}&\rule{0.5cm}{0pt}&\rule{1cm}{0pt}&\rule{1cm}{0pt}&\rule{0.5cm}{0pt}
            \\[-\arraystretch\normalbaselineskip]
            & {$L/a$}& {$1/aT$} & {$L_0/a$} & {$a\,\mathrm{(fm)}$} & {$6/g_0^2$} & {$N_\mathrm{conf}$}\\
            \midrule
            F7 & 48 & 12 & 96   & 0.0658 & 5.3      & 482  \\
            O7 & 64 & 16 & 128  & 0.049  & 5.5      & 305  \\
            W7 & 80 & 20 & {--} & 0.039  & 5.685727 & 1566 \\
            X7 & 96 & 24 & {--} & 0.033  & 5.82716  & 511  \\
            \bottomrule
        \end{tabular}
    \end{center}
    \caption{The lattice parameters for the numerical study with $\Nf=2$
        $\mathrm O(a)$-improved Wilson fermions. The vacuum ensembles are from
        the CLS collaboration and the tuning to the line of constant physics
        for the fine thermal ensembles was performed in
        ref.~\cite{Steinberg:2021bgr}. The pseudoscalar mass in the vacuum is
        approximately $m_\pi\approx270\,\mathrm{MeV}$ and the temperature is
        $T=254\,\mathrm{MeV}$.
        Note the large lattice sizes and small lattices spacings available for
        the thermal case.}
    \label{tab:nf2}
\end{table}

Given the encouraging results from the leading-order theory, we investigated
the improvement strategy with $\Nf=2$ non-perturbatively $\mathrm
O(a)$-improved Wilson fermions in the sea.
Details of the ensembles used to compute the thermal observable with lattice
spacings down to $a_\mathrm{th}\sim0.03\,\mathrm{fm}$, and down to
$a\sim0.05\,\mathrm{fm}$ for the vacuum case.
The line of constant physics set by the quark mass and $L$ is fixed going
toward the continuum limit in both the vacuum and thermal cases. 
{Although the current is not on-shell $\mathrm O(a)$-improved, we expect, due
to the lack of chiral-symmetry breaking at short distances} (and high
temperatures~\cite{DallaBrida:2020gux}), {that the $\mathrm{O}(a)$ cutoff
effects will be suppressed and proportional to the quark mass, given that
off-shell contributions contribute at higher order.}
We use the results of the previous section to also implement the subtraction
of the lattice artifacts at leading order in perturbation theory for the
thermal observable according to
\begin{align}
    \mathring{\mathcal I}^\mathrm{th}(t,a) &= \mathcal I^\mathrm{th}(t,a)
    - \Big[\mathcal I^\mathrm{th}(t,a) - I^\mathrm{th}(t)\Big]_\mathrm{LO}.
    \label{eq:tli}
\end{align}

\begin{figure}[t]
    \hspace{-2em}
    \begin{center}
        \includegraphics[scale=0.65]{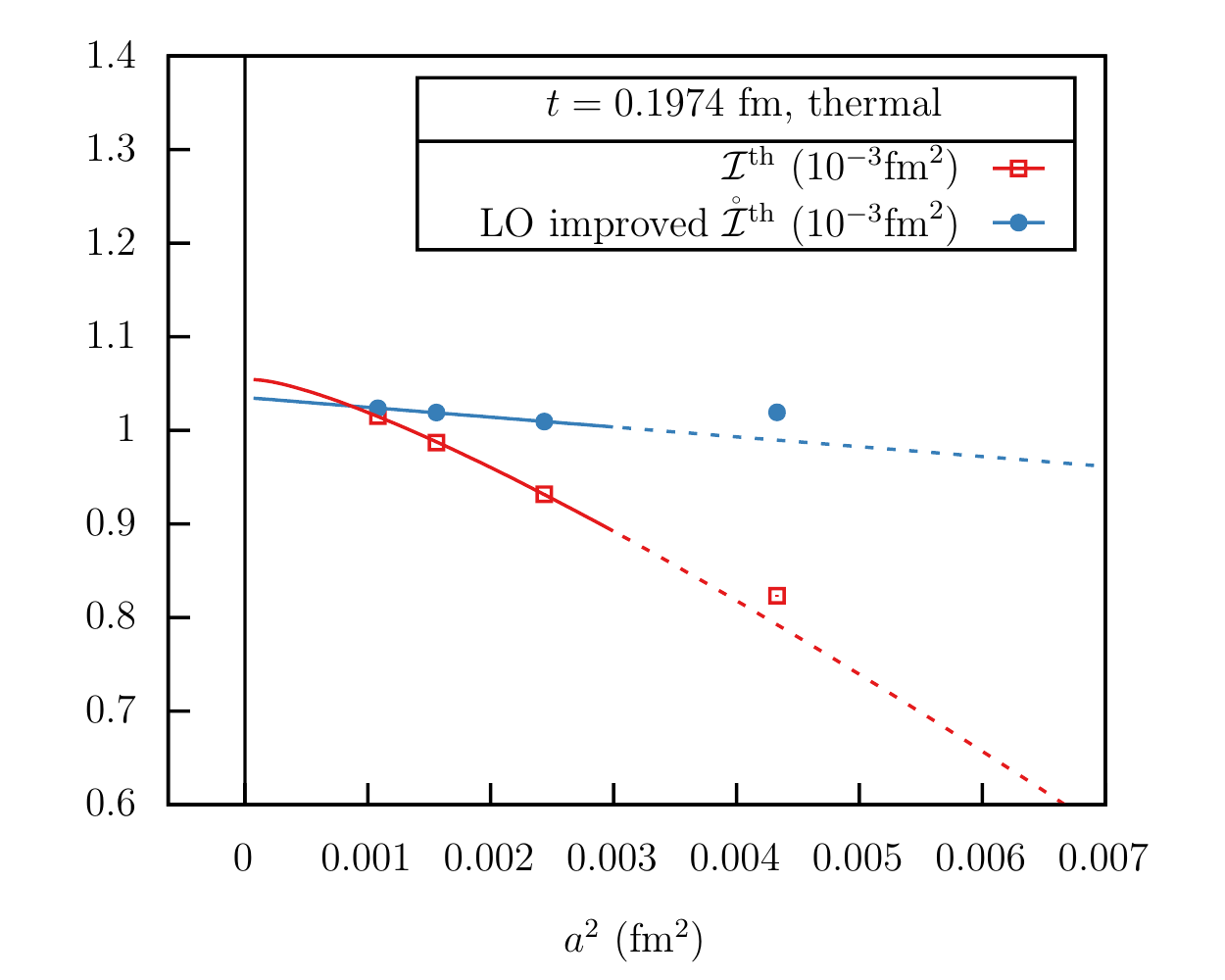}%
        \includegraphics[scale=0.65]{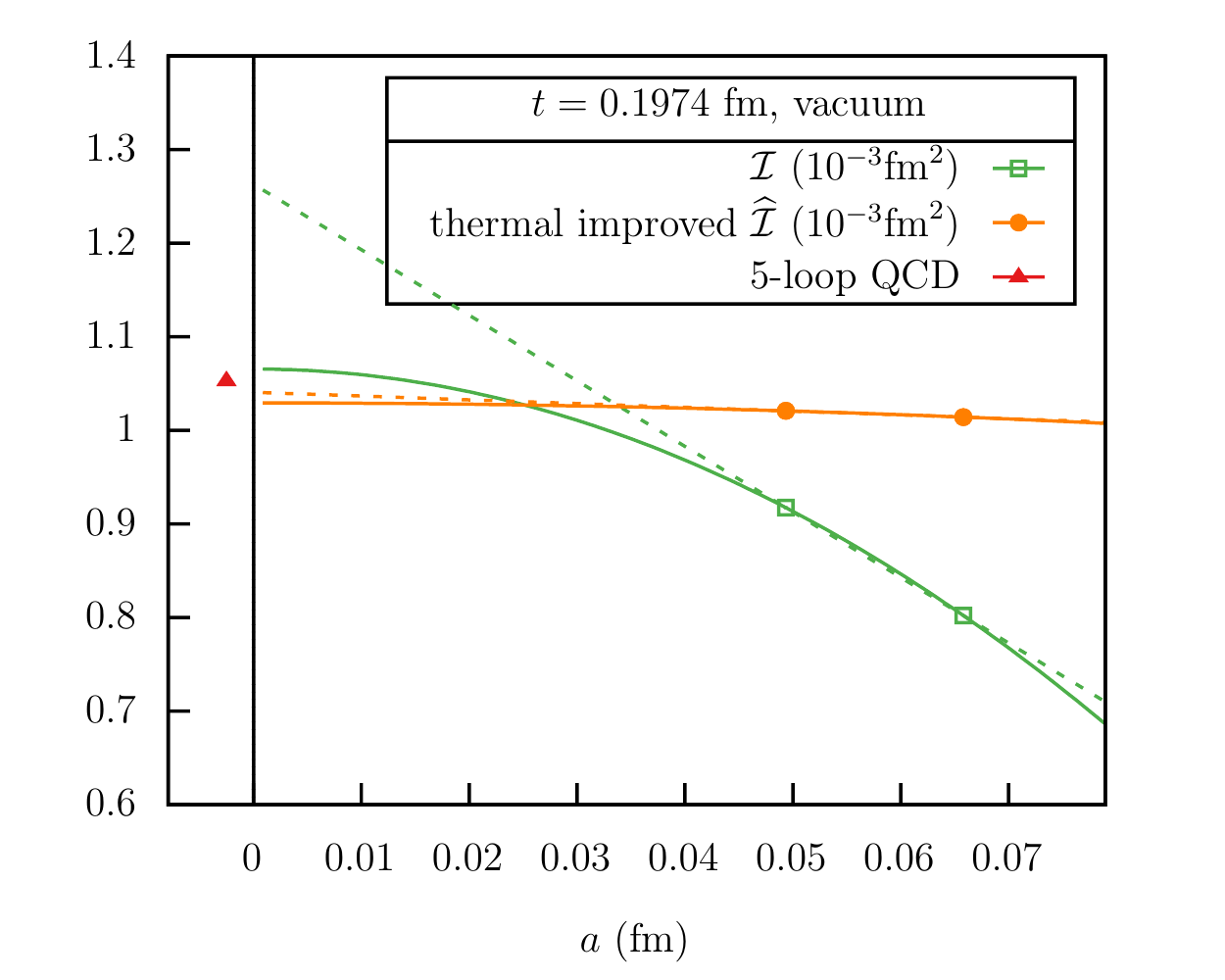}
    \end{center}
    \caption{The continuum limit for the thermal (left) and vacuum (right) observable
    with $\Nf=2$ $\mathrm O(a)$-improved Wilson fermions.
    In the thermal case (red), the Ansatz is quadratic in $a$ including the leading-order
    logarithmic artifact, while for the leading-order improved observable
    (blue) it is linear in $a^2$.
    In the vacuum, the case with (without) thermal subtraction is shown in
    orange (green).
    Ans\"atze linear in $a$ (dashed) and in $a^2$ (solid) are shown in both
    cases in the vacuum.}
    \label{fig:nf2}
\end{figure}

In figure~\ref{fig:nf2} (left), the addition of the two fine lattices for the
thermal observable clearly improves the robustness of the continuum limit,
for $t=0.1974\,\mathrm{fm}$.
We use a systematic error from half of the difference between the
two extrapolations depicted in the figure, which are explained in the caption.
The subtraction of the leading-order lattice artifacts (blue) is beneficial in
removing a large fraction of the lattice artifacts in the thermal
observable.
For the vacuum observable, the lattice artifacts which remain after
subtracting the thermal component (orange) are significantly smaller, as was
observed in the leading-order computation.
In contrast to the unimproved data (green), the resulting extrapolation is
under much better control and a much smaller estimate of the corresponding
systematic error is obtained.

For illustration, a final result can be obtained by combining the thermal
observable with the subtraction of leading-order lattice artifacts and the
mean of the linear and quadratic extrapolations for the remainder
\begin{align}
    I(t=0.1974\,\mathrm{fm}) = 1.035(9)_\mathrm{stat}(19)_\mathrm{cont}\times
    10^{-3}\mathrm{fm}^2,
    \label{eq:final}
\end{align}
while the systematics are added in quadrature.
We find satisfactory agreement (within the unaccounted mass and
non-perturbative effects) with the result from perturbation theory using the
known five-loop spectral density, from which we obtain
$I_\mathrm{pert}(t=0.1974\,\mathrm{fm})=1.059_{-6}^{1}\times10^{-3}\mathrm{fm}^2$,
where the asymmetric errors are due to the uncertainty in the $\Nf=2$
$\Lambda_{\overline{\mathrm{MS}}}$.

\section{The (discrete) Adler function at any scale}
\label{sec:hvp}

The analysis of the temperature effects on the short-distance current
correlator leads to a practical strategy for computing non-perturbatively the
Adler function, or its discrete analogue $\Delta_2(Q^2)=\Pi(Q^2)-\Pi(Q^2/4)$,
at arbitrarily high energies $Q^2$ while keeping cutoff effects under control.
The preceding analysis suggests that in fact a good precision at the percent
level can be obtained in a single-scale simulation for the thermal analogue
of $\Delta_2(Q^2)$ with $T=Q/8\pi$.
It is seemingly impossible to satisfy simultaneously the conditions
\begin{align}
    T\lesssim \Lambda\qquad\textrm{and}\qquad Q\ll a^{-1}
    \label{eq:twohier}
\end{align}
for arbitrarily large $Q$ to obtain a vacuum estimate.
However, for large $Q^2$, the temperature effects are known to be of order
$(\pi T/Q)^4$, which suggests that an accurate estimate of the vacuum
correction would be obtained by including the difference to $T/2$ as
\begin{align}
    \Delta_2(Q^2) &\approx \Delta_2(Q^2;T) - \Large[\Delta_2(Q^2;T) -
    \Delta_2(Q^2;T/2)\Large].
    \label{eq:discadler}
\end{align}
In this case only the double hierarchy
\begin{align}
    T\ll Q \ll a^{-1}
    \label{eq:double}
\end{align}
needs to be satisfied as the vacuum scale has been removed and the temperature
$T$ has been linked to the scale $Q$.
This allows any range of energies to be explored with this given level of
precision.%
This strategy has also been validated at leading order in perturbation
theory~\cite{Ce:2021xgd}, where indeed the omitted correction from $T/2$ to
the vacuum was observed to be a relative effect at the per-mille level.
In the Schwinger model a related strategy has also been presented at this
conference~\cite{Frech:2021bmw}.

\section{Conclusions}
\label{sec:concs}
The study of the temperature effects on the short-distance electromagnetic
current correlator has led to new insights to control the cutoff effects on
vacuum quantities like the Adler function at large energies, or the
short-distance HVP contribution to the muon anomalous magnetic moment.
These effects were investigated using the operator-product expansion at finite
temperature which has also been worked out at next-to-leading
order~\cite{Ce:2020wgg}.
A full computation of the lattice correlator at leading order in perturbation
theory with massless Wilson fermions allowed the asymptotics to be
cross-checked, and to study the proposal to use a thermal subtraction to
improve the continuum limit.
A resulting observation was that, for any discretization, a
logarithmically-enhanced $\mathrm{O}(a^2)$ lattice artifact is present in
these observables, which highlights that the continuum limit must be taken
with care.
The strategy was implemented for $\Nf=2$ non-perturbatively
$\mathrm{O}(a)$-improved Wilson fermions for the HVP contribution to muon
anomaly up to $x_0\approx0.2\,\mathrm{fm}$ and utilizing a thermal ensemble with
a temperature of $T\approx 250\,\mathrm{MeV}$.
The use of the thermal subtraction gave a clear advantage, while the
subtraction of the leading-order lattice artifacts was beneficial where small
lattice spacings were available.
Finally, a step-scaling strategy was outlined to compute a discrete analogue of
the Adler function at arbitrarily large energies, which is relevant to
determine the hadronic contribution to the running of the electromagnetic
coupling at the $Z$-pole.

\section*{Acknowledgments}
This work was supported by the European Research Council (ERC) under the
European Union’s Horizon 2020 research and innovation program through Grant
Agreement No. 771971-SIMDAMA, as well as by the Deutsche Forschungsgemeinschaft
(DFG, German Research Foundation) through the Cluster of Excellence ``Precision
Physics, Fundamental Interactions and Structure of Matter'' (PRISMA+ EXC
2118/1) funded by the DFG within the German Excellence strategy (Project ID
39083149).
The work of M.C. is supported by the European Union’s Horizon 2020 research and
innovation program under the Marie Skłodowska-Curie Grant Agreement No.
843134-multiQCD.
T.H. is supported by UK STFC CG ST/P000630/1.
The generation of gauge configurations as well as the computation of
correlators was performed on the Clover and Himster2 platforms at
Helmholtz-Institut Mainz and on Mogon II at Johannes Gutenberg University
Mainz.
The authors gratefully acknowledge the Gauss Centre for Supercomputing e.V.
(www.gauss-centre.eu) for funding project IMAMOM by providing computing time
through the John von Neumann Institute for Computing (NIC) on the GCS
Supercomputer JUWELS~\cite{JUWELS} at J\"ulich Supercomputing Centre (JSC).
Our programs use the QDP++ library~\cite{Edwards:2004sx} and deflated SAP+GCR
solver from the openQCD package~\cite{Luscher:2012av}.
We are grateful to our colleagues in the CLS initiative for sharing the gauge
field configurations on which this work is partially based.

\printbibliography
%

\end{document}